\documentclass[RNAAS]{aastex62}

\begin{document}

\title{High-Resolution Radio Image of a Candidate Radio Galaxy at $z=5.72$}

\correspondingauthor{Krisztina \'Eva Gab\'anyi}
\email{krisztina.g@gmail.com}

\author{Krisztina \'Eva Gab\'anyi}
\altaffiliation{}
\affiliation{MTA-ELTE Extragalactic Astrophysics Research Group, P\'azm\'any s\'et\'any 1/A, H-1117 Budapest, Hungary}
\affiliation{Konkoly Observatory, MTA CSFK, Konkoly Thege \'ut 15-17, H-1121 Budapest, Hungary}

\author{S\'andor Frey}
\affiliation{Konkoly Observatory, MTA CSFK, Konkoly Thege \'ut 15-17, H-1121 Budapest, Hungary}

\author{Leonid I. Gurvits}
\affiliation{Joint Institute for VLBI ERIC, Oude Hoogeveensedijk 4, 7991 PD Dwingeloo, the Netherlands}
\affiliation{Department of Astrodynamics and Space Missions, Delft University of Technology, Kluyverweg 1, NL-2629 HS Delft, the Netherlands}

\author{Zsolt Paragi}
\affiliation{Joint Institute for VLBI ERIC, Oude Hoogeveensedijk 4, 7991 PD Dwingeloo, the Netherlands}

\author{Krisztina Perger}
\affiliation{Department of Astronomy, E\"otv\"os Lor\'and University, P\'azm\'any s\'et\'any 1/A, H-1117 Budapest, Hungary}
\affiliation{Konkoly Observatory, MTA CSFK, Konkoly Thege \'ut 15-17, H-1121 Budapest, Hungary}



\section{} 

\cite{disc} reported the discovery of a possible radio galaxy at a redshift of $z=5.72$, based on the detection of a single Ly$\alpha$ emission line. If it is indeed a radio galaxy, this would be the most distant known object of this type. 
The authors collected a sample of ultra-steep spectrum sources, with the spectral index $\alpha < -1.3$ ($S \sim \nu^{\alpha}$, where $\nu$ is the frequency and $S$ is the flux density measured between $150$ MHz and $1.4$ GHz) and with compact radio morphologies using the TIFR GMRT Sky Survey Alternative Data Release \citep{Intema}, the Faint Images of the Radio Sky at Twenty-Centimeters \citep{first} and the 1.4-GHz NRAO VLA Sky Survey \citep{nvss}. Only sources which were not detected in various optical (SDSS DR12, \citealt{sdss}; PAN-STARRS1, \citealt{panstarrs}) and infrared surveys (AllWISE, \citealt{wise}; UKIDSS, \citealt{ukids}) were further imaged with the Karl G. Jansky Very Large Array (VLA) at $1.4$ GHz in its most extended A configuration. TGSS1530 (hereafter J1530$+$1049) was one of their brightest sources detected with a flux density of $S=7.5\pm 0.1$ mJy. It was unresolved in the VLA-A observation and its spectral index is $-1.4\pm 0.1$. 

\begin{figure}[h!]
\begin{center}
\includegraphics[width=0.75\columnwidth,bb=130 40 810 1200 clip]{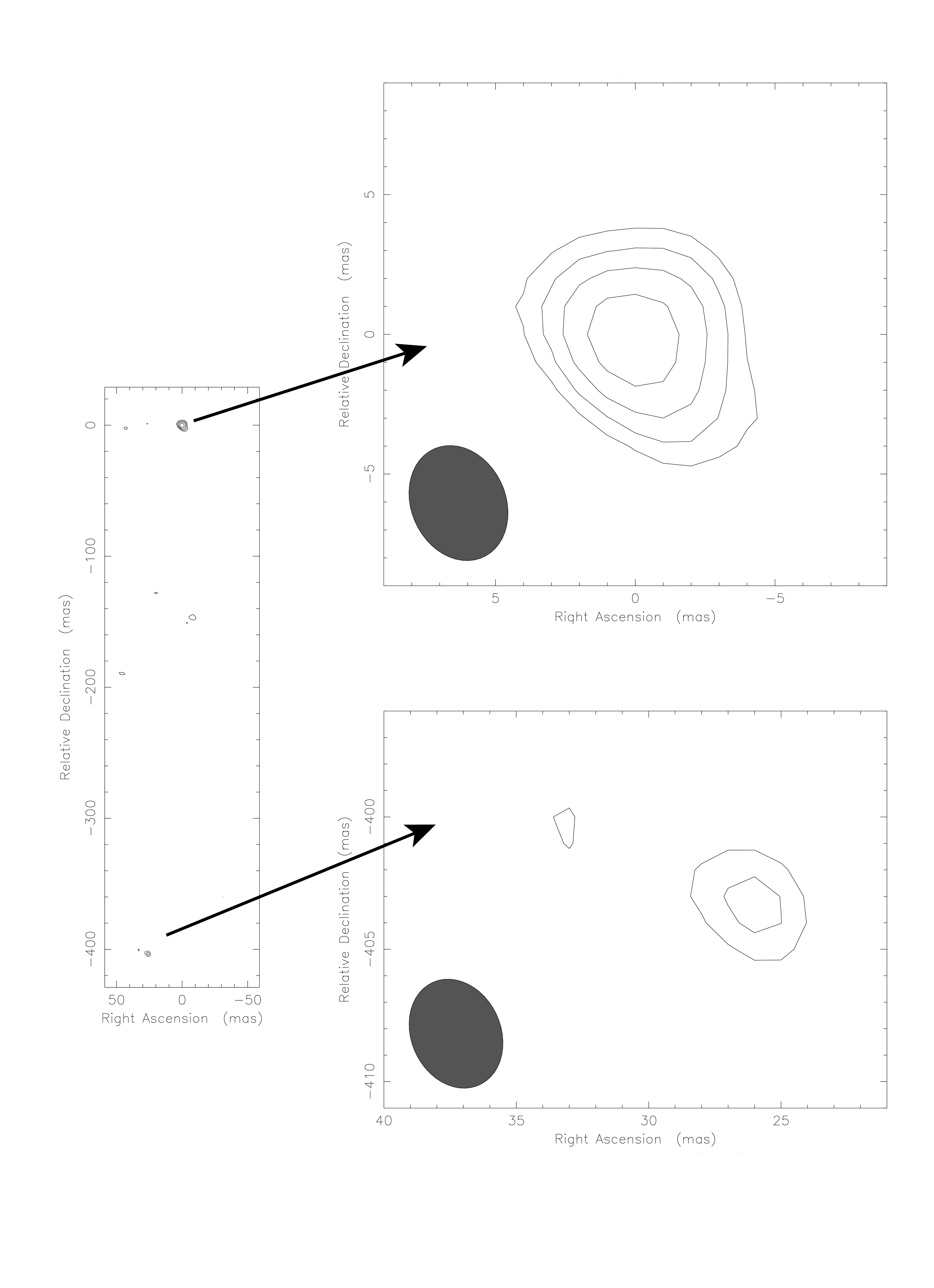}
\caption{1.7-GHz EVN radio image of J1530$+$1049. Peak intensity is $0.5 \mathrm{\,mJy\,beam}^{-1}$, the lowest contour is at  $0.1 \mathrm{\,mJy\,beam}^{-1}$ ($3\sigma$ noise level), further contour levels increase by a factor of $\sqrt{2}$. The beam size is $4.3$\,mas $\times 3.4$\,mas at position angle $24\degr$. \label{fig:1}}
\end{center}
\end{figure}

We observed J1530$+$1049 with the European Very Long Baseline Interferometry (VLBI) Network (EVN) at $1.7$\,GHz on 2018 Sep 19. The following radio telescopes provided data: a single antenna of the Westerbork Synthesis Radio Telescope (the Netherlands), Effelsberg (Germany), Medicina (Italy), Onsala (Sweden), Tianma (China), Toru\'n (Poland), Hartebeesthoek (South Africa), and Sardinia (Italy). 
Eight $16$-MHz wide intermediate frequency channels were used in left and right circular polarizations. 
The observation was conducted in phase-reference mode \citep{phase-ref}. The target and the phase-reference calibrator (J1525$+$1107, its coordinates are known within an accuracy of $0.2$\,mas\footnote{{\url hpiers.obspm.fr/icrs-pc/newwww/icrf/}, Charlot et al., in prep.}) were observed alternately to facilitate the detection of the faint target and its precise relative astrometry.
On-source time was $1.3$\,h. For the details of data reduction we refer to \cite{reduction}. 
We detected two faint radio features in J1530$+$1049 with a separation of $\sim 400$\,mas (Fig. \ref{fig:1}), corresponding to $\sim 2.5$\,kpc at $z=5.72$ (assuming a flat $\Lambda$CDM cosmological model with $H_0=70$\,km\,s$^{-1}$\,Mpc$^{-1}$, $\Omega_\mathrm{m}=0.27$). The position of the brighter northern feature is right ascension $15^\mathrm{h} 30^\mathrm{m} 49\fs8903$ and declination $+10\degr 49' 31\farcs175$ with $1$\,mas estimated accuracy. The sum of the flux densities of the two components is $1.7\pm0.2$\,mJy. Even taking into account its steep spectrum, the EVN observations recovered only a fraction of the flux density extrapolated from the VLA value. While this can be related to variability since the radio observations were not simultaneous, it is more probable that the missing flux density is in sub-arcsec structure compact on the VLA scale but resolved out by the EVN.

The radio power calculated from the VLA flux density ($\sim 10^{28} \mathrm{\,W\,Hz}^{-1}$, \citealt{disc}) and the projected source size derived from our EVN data place J1530$+$1049 among the medium-sized symmetric objects (MSOs). These are young counterparts of radio galaxies in the evolutionary diagram of \cite{mso}. This is consistent with a radio galaxy in an early phase of its evolution as proposed by \cite{disc}. Note that \cite{momjian} recently imaged with VLBI a radio quasar at $z=5.84$ that possibly shows MSO structure. 

From our single-frequency radio image of J1530$+$1049, it is not possible to decide whether any of the components detected is a flat-spectrum radio core or both are steeper-spectrum hot spots where the jets interact with the dense interstellar medium of the host galaxy. Multi-frequency interferometric observations with $\sim 0.1$ arcsec resolution could answer this question.



\acknowledgments

K\'EG acknowledges the J\'anos Bolyai Research Scholarship of the Hungarian Academy of Sciences. This work was supported by the NKFIH-OTKA NN110333 grant. The EVN is a joint facility of independent European, African, Asian, and North American radio astronomy institutes. Scientific results from data presented in this publication are derived from the following EVN project code: RSG11. This project has received funding from the European Union's Horizon 2020 research and innovation programme under grant agreement No 730562 [RadioNet].

\end{document}